\documentstyle[12pt,amssymbols]{article}
\newcommand{\be}{\begin{equation}}
\newcommand{\ee}{\end{equation}}
\newcommand{\bea}{\begin{eqnarray}}
\newcommand{\eea}{\end{eqnarray}}
\newcommand{\brr}{\begin{array}}
\newcommand{\err}{\end{array}}
\newcommand{\bc}{\begin{center}}
\newcommand{\ec}{\end{center}}
\newcommand{\nn}{\nonumber}
\newcommand{\sss}{\scriptscriptstyle}

\newcommand{\ep}{\frac{1}{\epsilon}}

\newcommand{\epuno}{\epsilon}
\newcommand{\epdue}{\epsilon^{ 2}}

\newcommand{\Z}{\hat{     Z}}

\newcommand{\LL}{\gamma^{\mu}_{ L}\otimes\gamma_{\mu L}}
\newcommand{\LR}{\gamma^{\mu}_{ L}\otimes\gamma_{\mu R}}

\newcommand{\alphas}{\alpha_{ s}}

\newcommand{\LV}{\gamma^{\mu}_{ L}\otimes\gamma_{\mu}}

\def\tsp{\rule{0pt}{2.1ex}}
\def\tsq{\rule{0pt}{2.5ex}}
\newcommand{\prd}{Phys. Rev. \underline}
\newcommand{\pl}{Phys. Lett. \underline}
\newcommand{\prl}{Phys. Rev. Lett. \underline}
\newcommand{\np}{Nucl. Phys. \underline}
\newcommand{\qslash}{/ \mskip-8mu q}
\newcommand{\pslash}{/ \mskip-10mu p}
\begin{document}
\setcounter{page}{1}
\pagestyle{empty}
\begin{flushright}
ROME prep. 973/93\\
INFN-ISS 93/6\\
ULB-TH 15/93\\
November 1993\\
\end{flushright}
\vskip 1cm
\centerline{\bf{Leading Order QCD Corrections to $b
\rightarrow s \, \gamma$ }}
\centerline{\bf{and $b \rightarrow s \, g$
Decays in Three Regularization Schemes}}
\vskip 1.5cm
\centerline{\bf{M. Ciuchini$^{a,b}$,
E. Franco$^b$, L. Reina$^c$ and L. Silvestrini$^b$ }}
\vskip .5cm
\centerline{$^a$ INFN, Sezione Sanit\`a, V.le Regina Elena 299,
00161 Roma, Italy. }
\centerline{$^b$ Dip. di Fisica,
Universit\`a degli Studi di Roma ``La Sapienza" and}
\centerline{INFN, Sezione di Roma, P.le A. Moro 2, 00185 Roma, Italy. }
\centerline{$^c$ Service de Physique Th\'eorique\footnote{Chercheur
IISN}, Universit\'e Libre de Bruxelles,}
\centerline{Boulevard du Triomphe, CP 225 B-1050 Brussels, Belgium.}
\vskip 2cm
\begin{abstract}
We discuss in detail the calculation of the leading order QCD corrections
to the Effective Hamiltonian which governs $b \rightarrow s\, \gamma$ and
$b \rightarrow s \, g$ transitions in three different regularization
schemes (HV, NDR and DRED). We show that intermediate stages of the
calculation do depend on the regularization, but the same scheme
independent coefficients can be obtained in all the considered schemes.
A detailed discussion of the results already present in the literature is
also given.
\end{abstract}
\date{}
\newpage
\setcounter{page}{1}
\pagestyle{plain}

\section{Introduction}
\label{sec:intro}

Leading order (LO) QCD corrections to the Effective Hamiltonian
governing the radiative decays of the $B$ meson have been calculated by many
authors in the last five years \cite{gsw1}-\cite{cfmrs1}. In fact they turn out
to be large and important for the $B$ phenomenology. In spite of this effort,
different results are still present in the recent literature on this subject,
even if the residual differences are numerically small and have essentially
no phenomenological relevance. The aim of this paper is to help clarifying
the origin of these differences by giving full details on our calculations
in three different regularization schemes.

Let us briefly recall how the calculation of LO corrections to
the Effective Hamiltonian for radiative $B$ decays has been developed
in the recent years.
The original calculations, refs. \cite{gsw1} and \cite{gosn1}, were performed
in the Naive Dimensional Regularization scheme (NDR) and in the Dimensional
REDuction scheme (DRED \cite{sieg}) respectively and used a reduced set of
operators for the Effective Hamiltonian. They disagree on the results for the
anomalous dimension matrix, that was believed to be regularization scheme
independent at the leading order.

Later NDR result was confirmed \cite{ccrv} and this led the
authors of ref. \cite{gosn3} to cast doubts on the reliability of the DRED
scheme.
Two years ago a first paper \cite{mis1} appeared where the complete
LO correction was calculated in the NDR scheme using the full set of
operators, then other similar calculations followed \cite{mis2}-\cite{yao1}.
Refs. \cite{mis2} and \cite{yao1} confirm the result of ref.
\cite{gsw1} for the anomalous dimension matrix in the ``reduced'' basis.
However they disagree on some new matrix elements in the ``full'' basis.

Recently the question of the scheme independence of the Effective Hamiltonian
for B meson radiative decays has been clarified \cite{cfmrs1}. The calculation
of the complete LO corrections has been performed in NDR and in the
't Hooft-Veltman scheme (HV \cite{hv}) and it has been shown that the final
results for the Wilson coefficients are regularization scheme independent
provided one takes properly into
account the scheme dependence of the one and two loop Feynman diagrams.
Calculations which use the reduced set of operators such as those of refs.
\cite{gsw1,gosn1,ccrv} have been demonstrated to give results which depend on
the regularization scheme. Unfortunately the new NDR anomalous dimension matrix
obtained in ref. \cite{cfmrs1} differs from both the results of refs.
\cite{mis2,yao1}.

In this paper we provide full details on the calculation presented
in ref. \cite{cfmrs1}, including tables with the pole coefficients for all the
diagrams. We also present the calculation in the DRED scheme.
We show how the same regularization independent results can be obtained in DRED
as well as in the other two schemes, thus extending the result of
ref. \cite{mis3}. We also discuss why the conclusions of ref. \cite{gosn3} on
the
reliability of the DRED scheme are wrong. Finally we compare our results with
the most recent ones \cite{mis2,yao1} and comment on the
differences among the results, reported in refs. \cite{mis3,yao2}.

The paper is organized as follows.
In Sec. \ref{sec:heff} we recall the Effective Hamiltonian for
$b \rightarrow s\, \gamma$ decays and the renormalization group equations
(RGE) governing the evolution of the Wilson coefficients. The full operator
basis is given and compared with the reduced one. The initial conditions for
the coefficients are also given.
In Sec. \ref{sec:solut} the explicit solution of the RGE, showing the scheme
independence of the coefficients, is given.
In Sec. \ref{sec:schemes} we summarize the definitions of NDR, HV and DRED
regularization schemes, discussing how they are implemented in our
calculations.
In Sec. \ref{sec:diag} the contributions of one loop diagrams to anomalous
dimension, operator matrix elements and counter-terms are considered and
diagram by diagram results are given. The same is done for the contributions
of the two loop diagrams to the anomalous dimension. We discuss the relations
among one and two loop diagrams induced by the scheme independence of the
final result and briefly report other checks done on our calculation.
In Sec. \ref{sec:dimanom} the anomalous dimension matrices in the different
schemes are given.
Finally in Sec. \ref{sec:status} we critically compare our results with the
other ones present in the literature.

\section {Effective Hamiltonian: Initial Conditions and Evolution for the
Coefficients}
\label{sec:heff}

The effective Hamiltonian for $b \rightarrow s \, \gamma$
($b \rightarrow s\, g$) transitions is given by
\be
 H_{eff}= -V_{tb} V^*_{ts} \frac {G_F} {\sqrt{2}}
\sum_{i=1}^{8} Q_{ i}(\mu) C_{ i}(\mu) \sim  \vec Q^T(\mu) \, \vec C(\mu)
\label{eh} \ee
where $V_{ij}$ are the elements of the CKM\cite{cab,km} quark mixing matrix.
We use the following operator basis $\vec Q$
\bea
Q_{ 1}&=&({\bar s}_{\alpha}c_{\beta})_{ (V-A)}
    ({\bar c}_{\beta}b_{\alpha})_{ (V-A)}
   \nn\\
Q_{ 2}&=&({\bar s}_{\alpha}c_{\alpha})_{ (V-A)}
    ({\bar c}_{\beta}b_{\beta})_{ (V-A)}
\nn \\
Q_{ 3,5} &=& ({\bar s}_{\alpha}b_{\alpha})_{ (V-A)}
    \sum_{q=u,d,s,\cdots}({\bar q}_{\beta}q_{\beta})_{ (V\mp A)}
\nn \\
Q_{ 4,6} &=& ({\bar s}_{\alpha}b_{\beta})_{ (V-A)}
    \sum_{q=u,d,s,\cdots}({\bar q}_{\beta}q_{\alpha})_{ (V\mp A)}
\nn \\
Q_{ 7} &=& \frac{Q_de}{16\pi^2}m_b{\bar s}_{\alpha}\sigma^{\mu\nu}_{(V+A)}
           b_{\alpha}F_{\mu\nu}
\nn \\
Q_{ 8} &=& \frac{g}{16\pi^2}m_b{\bar s}_{\alpha}\sigma^{\mu\nu}_{(V+A)}
           t^A_{\alpha\beta}b_{\beta}G^A_{\mu\nu}.
\label{basis}
\eea
Here $(V\mp A)$ indicate the chiral structure, $\alpha$ and $\beta$ are colour
indices, $m_b$ is the $b$ quark mass, $Q_d=-\frac{1}{3}$ is the electric charge
of the down-type quarks and $g$ ($e$) is the strong (electro-magnetic)
coupling.
The colour matrices normalization is $ Tr(t^A t^B)= \delta^{AB}/2$.

The choice of the operator basis deserves some comments. In fact the basis
in eq. (\ref{basis}) is obtained by using the equations of motion of the
external fields. While this procedure has been criticized in the past, a
recent paper \cite{simma} shows that it can be safely used.
This basis is often called the ``complete'' basis, to be compared with the
``reduced'' one of refs. \cite{gsw1,gosn1,ccrv}. Using the ``reduced'' basis,
one neglects the contributions of the operators $O_3,\dots,O_6$.
Also retaining the penguin operator\footnote{In the ``complete'' basis this
operator can be eliminated via the equations of motion.}
$O_P = \bar s \gamma^{\mu}_L D^\nu  t^A G^A_{\mu\nu} b$
as in ref. \cite{ccrv}, one actually does not fully consider the
contribution of $O_3,\dots,O_6$. Unfortunately just the insertion
of $O_5,\,O_6$ in the one loop diagrams gives a scheme
dependent contribution, which does not vanish in the NDR scheme.
Hence NDR calculations using the reduced basis give
a regularization dependent result, see ref. \cite{cfmrs1} and Secs.
\ref{sec:solut}, \ref{sec:diag}, \ref{sec:status} below.

The coefficients $\vec C(\mu)$ of eq. (\ref{eh}) obey the renormalization
group equations
\be
\left( - \frac {\partial} {\partial t} + \beta ( \alphas )
\frac {\partial} {\partial \alphas} - \frac {\hat \gamma^T ( \alphas ) }{2}
 \right) \vec C(t, \alphas(t)) =0 \label{rge}, \ee
where $t=\ln(M_W^2/\mu^2)$ and $\alpha_s=g^2/4\pi$.
The factor of $2$ in eq. (\ref{rge}) normalizes the anomalous dimension
matrix as in ref. \cite{cfmr2}. $\hat\gamma$ includes the contribution due
to the renormalization of $m_b$, the gluon field and the strong coupling
constant $g$, see e.g. refs. \cite{ccrv,svz} (see also Sec.
\ref{sec:dimanom}).

The initial conditions for the coefficients can be easily found by matching
the effective theory with the ``full'' theory at the scale $M_W$. They are
given by \cite{desnaz}
\bea
C_1(M_W)&=&0 \nn\\
C_2(M_W)&=&1 \nn\\
C_{3}(M_W),\dots,C_{6}(M_W)&=&0 \nn\\
C_{ 7} (M_{ W}) &=&  -3\frac{3
x^{ 3} - 2 x^{ 2}}{2(1 -
x)^{ 4}}\ln {x} - \frac{8 x^{ 3}+ 5
x^{ 2} - 7 x}{4 (1 - x)^{ 3}} \nn\\
C_{ 8} (M_{ W}) &=&
-\frac{3 x^{ 2}}{2 (1-x)^{
4}}\ln {x} + \frac{x^{ 3}-5 x^{
2}- 2 x}{4 (1 - x )^{ 3}},
\label{inicoef}
\eea
where $x=m_t^2/M_W^2$.

\section {Regularization Scheme Independence of the Effective Hamiltonian}
\label{sec:solut}
In this section the explicit solution of eq. (\ref{rge}) for the
coefficients $\vec C(\mu)$ is given and the regularization scheme
independence of the Effective Hamiltonian is discussed, following ref.
\cite{cfmrs1}.
The solution in eq. (\ref{evolu2}), as well as the relation among matrices
in different schemes in eq. (\ref{check2}), rely on the peculiar structure
of the anomalous dimension matrix, which is
\be
\hat{\gamma}=\frac{\alphas}{4\pi}
\left( \begin{array}{ccc}
\hat\gamma_r & \vec\beta_7 & \vec\beta_8\\
\vec 0^{\scriptscriptstyle T} & \gamma_{77} & 0 \\
\vec 0^{\scriptscriptstyle T} & \gamma_{87} & \gamma_{88}
\end{array} \right).
\label{matform}
\ee
This ``almost'' triangular form is obtained because the magnetic operators
$Q_7$, $Q_8$ do not mix with the 4-fermion operators $Q_1,\dots,Q_6$.
Then it is convenient to
introduce the reduced $6\times 6$ matrix $\hat\gamma_r$ which mixes
$Q_1,\dots, Q_6$ among themselves and two 6-component column vectors
$\vec\beta_7=\left( \gamma_{17},\gamma_{27},\dots,\gamma_{67} \right)$
and $\vec\beta_8=\left(\gamma_{18},\gamma_{28},\dots,\gamma_{68}\right)$,
which account for the two-loop mixing of the 4-fermion operators with
the magnetic ones.

A very peculiar characteristic of this calculation is that at the
LO of the perturbative expansion both one-loop and two-loop diagrams
contribute. The mixing of 4-fermion operators among themselves
($\hat\gamma_r$) as well as the mixing of magnetic-type operators
among themselves ($\gamma_{77}$, $\gamma_{87}$, $\gamma_{88}$) are
one-loop effects (at LO), while the mixing between 4-fermion and
magnetic-type operators ($\vec\beta_7$, $\vec\beta_8$) is a two-loop
effect (always at LO).

In terms of these quantities the RGE are given by
\bea
2\mu^2\frac {d}{d\mu^2}\vec C_r(\mu) &=& \frac{\alphas}{4\pi}
   \hat\gamma^T_r\vec C_r(\mu)
\nn \\
2\mu^2\frac {d}{d\mu^2} C_7(\mu) &=& \frac{\alphas}{4\pi}\left(
   \vec\beta_7\cdot\vec C_r(\mu)+\gamma_{77}C_7(\mu)+\gamma_{87}C_8(\mu)\right)
\nn \\
2\mu^2\frac {d}{d\mu^2} C_8(\mu) &=& \frac{\alphas}{4\pi}\left(
   \vec\beta_8\cdot\vec C_r(\mu)+\gamma_{88}C_8(\mu)\right),
\label{evolu1}
\eea
where $\vec C_r(\mu)=\left(C_1(\mu),\dots,C_6(\mu)\right)$,
$\alphas=\alphas(\mu)$ and
$\mu^2 {d}/{d\mu^2}=\mu^2 {\partial}/{\partial\mu^2}+\beta
(\alpha_s){\partial}/
{\partial\alpha_s}$.
Diagonalizing the submatrix which mixes the magnetic operators, one obtains
\bea
2\mu^2\frac {d}{d\mu^2}\vec C_r(\mu) &=& \frac{\alphas}{4\pi}
   \hat\gamma^T_r\vec C_r(\mu)
\nn \\
2\mu^2\frac {d}{d\mu^2} v_7(\mu) &=& \frac{\alphas}{4\pi}\gamma_{77}v_7(\mu)
\nn \\
2\mu^2\frac {d}{d\mu^2} v_8(\mu) &=& \frac{\alphas}{4\pi}\gamma_{88}v_8(\mu),
\label{evolu2}
\eea
where
\bea
v_7(\mu) &=& C_7(\mu)+\vec\alpha_7\cdot\vec C_r(\mu)+\frac{\gamma_{87}}
 {\gamma_{77}-\gamma_{88}}C_8(\mu)\nn\\
v_8(\mu) &=& C_8(\mu)+\vec\alpha_8\cdot\vec C_r(\mu)
\label{eigenvec}
\eea
with
\bea
\vec\alpha_7 &=& \left(\gamma_{77}\hat 1-\hat\gamma_r\right)^{-1}\left[
  \vec\beta_7+\frac{\gamma_{87}}{\gamma_{77}-\gamma_{88}}\vec\beta_8\right]
\nn\\
\vec\alpha_8 &=& \left(\gamma_{88}\hat 1-\hat\gamma_r\right)^{-1}
   \vec\beta_8.
\label{alphas}
\eea

The rhs of eqs. (\ref{evolu2}) involves only one loop quantities, so that
the solutions
$\vec C_r(\mu)$, $v_7(\mu)$ and $v_8(\mu)$ are independent of the
regularization
scheme.
However the expression of both $v_7(\mu)$ and $v_8(\mu)$, eq.
(\ref{eigenvec}), contains both scheme-dependent and -independent
quantities.
In particular $\vec C_r(\mu)$ is the LO solution of the RGE for the
4-fermion operators, which is known to be regularization scheme independent.
Indeed, $\vec C_r(\mu)$ is known up to next-to-leading order terms
\cite{cfmr2,bjlw1}, which are not to be considered in our LO calculation.
On the contrary $\vec\beta_7$ and $\vec\beta_8$, hence $\vec\alpha_7$,
$\vec\alpha_8$ and $C_7(\mu)$, $C_8(\mu)$, come from two loop diagrams (see
Sec.
\ref{sec:diag} below) and they do depend on regularization scheme, even if
they are LO results. Precisely this point, that was missed in refs.
\cite{gsw1}-\cite{mis2}, is responsible for the difference between
previous results in NDR and DRED schemes, as stressed in ref. \cite{cfmrs1}.

Moreover, already noted in ref. \cite{mis1}, the operators
$Q_5$, $Q_6$ have non zero matrix elements between the $b$ and
$s\gamma$ ($s\, g$) states, through the penguin
diagrams in fig. (\ref{fig:1loop_ct}) with massive $b$ loop propagators.
These one loop matrix elements and the tree level matrix elements of the
magnetic operators are of the same order in $\alphas$, so that the matrix
elements of the Effective Hamiltonian can be written as
\bea
<s\gamma|H_{eff}|b>&=&C_7(\mu)<s\gamma|Q_7(\mu)|b>+
C_5(\mu) <s\gamma|Q_5(\mu)|b> \nn\\
&&+ C_6(\mu) <s\gamma|Q_6(\mu)|b>\nn\\
&=&\tilde C_7(\mu)<s\gamma|Q_7(\mu)|b>\nn\\
&&\nn\\
<sg|H_{eff}|b>&=&C_8(\mu)<sg|Q_8(\mu)|b>+C_5(\mu) <sg|Q_5(\mu)|b> \nn\\
&=&\tilde C_8(\mu)<sg|Q_8(\mu)|b>,
\label{matele}
\eea
where the coefficients $\tilde C_7(\mu)$, $\tilde C_8(\mu)$ are defined as
\bea
\tilde C_7(\mu)=C_7(\mu)+\vec Z_7\cdot\vec C_r(\mu)\nn\\
\tilde C_8(\mu)=C_8(\mu)+\vec Z_8\cdot\vec C_r(\mu).
\label{cz}
\eea
Eqs. (\ref{matele}) and (\ref{cz}) can be seen as a finite
renormalization of the operators such that the matrix elements of
$Q_5$ and $Q_6$ vanish.
The vectors $\vec Z_7$ and $\vec Z_8$ can be considered as the effect of a
mixing of order $\alphas^0$ among $Q_5$, $Q_6$ and the magnetic operators.
They are calculated from the finite part of the penguin diagrams
in fig. (\ref{fig:1loop_ct}), thus they depend on the regularization scheme.
We find that they vanish in HV and DRED, while in NDR they are given by
\bea
\vec Z_7^{\scriptscriptstyle NDR}&=&\left(0,0,0,0,2,2N\right)\nn\\
\vec Z_8^{\scriptscriptstyle NDR}&=&\left(0,0,0,0,2,0\right).
\label{zv}
\eea

We now show how the scheme independence of the Effective Hamiltonian is
recovered. In terms of the operators renormalized at the scale $\mu$, the
Effective Hamiltonian is given by
\bea
H_{eff}&\sim&\vec Q_r^T(\mu)\cdot \vec C_r(\mu) \nn \\
&&+\left\{ v_7(\mu)+\left[\left( \hat\gamma_r-\gamma_{77}\hat 1\right)^{-1}
\left(\vec\beta_7+\frac{\gamma_{87}}{\gamma_{77}-\gamma_{88}}\vec\beta_8
\right)+\vec Z_7\right]\cdot\vec C_r(\mu) \right. \nn \\
&&+\left. \frac{\gamma_{87}}{\gamma_{88}-\gamma_{77}}\left[v_8(\mu)+
\left(\hat\gamma_r-\gamma_{88}\hat 1\right)^{-1}\vec\beta_8\cdot
\vec C_r(\mu)\right]\right\}Q_7(\mu) \nn \\
&&+\left\{v_8(\mu)+\left[
\left(\hat\gamma_r-\gamma_{88}\hat 1\right)^{-1}\vec\beta_8 + \vec Z_8
\right]\cdot\vec C_r(\mu)\right\}Q_8(\mu) \nn \\
\nn\\
&=&\vec Q_r^T(\mu)\cdot \vec C_r(\mu) \nn \\
&&+\left\{ v_7(\mu)+\vec\omega_7\cdot\vec C_r(\mu)+
\frac{\gamma_{87}}{\gamma_{88}-\gamma_{77}}\left[v_8(\mu)+
\vec\omega_8\cdot\vec C_r(\mu)\right]\right\}Q_7(\mu) \nn \\
&&+\left\{v_8(\mu)+\vec\omega_8\cdot\vec C_r(\mu)\right\}Q_8(\mu),
\label{heff}
\eea
where
\bea
\vec\omega_7&=&\left( \hat\gamma_r-\gamma_{77}\hat 1\right)^{-1}
\left(\vec\beta_7+\frac{\gamma_{87}}{\gamma_{77}-\gamma_{88}}\vec\beta_8\right)
+\vec Z_7+\frac{\gamma_{87}}{\gamma_{77}-\gamma_{88}}\vec Z_8 \nn \\
\vec\omega_8&=&\left(\hat\gamma_r-\gamma_{88}\hat 1\right)^{-1}\vec\beta_8
+ \vec Z_8.
\label{omega}
\eea
Thus the Effective Hamiltonian is scheme independent provided that
$\vec\omega_7$ and $\vec\omega_8$ do not change with the scheme. We will prove
in the following that this is indeed the case.

Now we discuss the change in the anomalous dimension matrix induced by
a change of the regularization scheme. The anomalous dimension matrix in a
given scheme ``$a$'' is defined as
\be
\hat\gamma^a=2\left(\hat Z^a(\mu)\right)^{-1}\mu^2\frac{d}{d\mu^2}\hat
Z^a(\mu),
\label{diman}
\ee
where $\hat Z^a$ is the matrix of the renormalization constants which gives
the renormalized operators in terms of the bare ones
\be
\vec  Q(\mu)=\left(\hat Z^a\right)^{-1}\vec Q_B.
\ee

Let us consider different renormalization schemes. The operators
renormalized through the usual $\overline{MS}$ subtraction change from
one scheme to another. In order to have the same renormalized
operators in all the schemes, one can define them using the $\overline{MS}$
procedure in a given scheme, then adopts a suitable non-minimal subtraction in
the other ones. So the $\overline{MS}$ renormalization constants in two
different schemes ``$a$'' and ``$b$'' can be related through the equation
\be
\hat Z^a=\hat Z^b\hat r,
\label{renmat}
\ee
where the matrix $\hat r$ accounts for the change of the subtraction procedure
in the scheme ``$b$'' necessary to define operators renormalized as in
the scheme ``$a$''. In our case the matrix $\hat r$ is expressed in terms
of the vectors $\vec Z$ introduced in eq. (\ref{cz}), as discussed in the
following.

The relation between the anomalous dimension
matrices in the schemes ``$a$'' and ``$b$'' is easily obtained from
eqs. (\ref{diman})-(\ref{renmat})
\be
\hat\gamma^a=\hat r^{-1}\hat\gamma^b\hat r.
\label{check}
\ee
This relation was already found in ref. \cite{bw1} and applied to the
next-to-leading calculation of the $\Delta S=1$ Effective Hamiltonian.
In that case $\hat r$ deviated from $\hat 1$ by terms of order $\alphas$.
Here we are not interested in such terms, but now, as a consequence of eqs.
(\ref{matele})-(\ref{zv}), $\hat r$ differs from the identity already
at order $\alphas^0$. However, due to the peculiar form of the anomalous
dimension matrix discussed above, the matrix $\hat r$ is simply
\be
\hat r=\left(\begin{array}{c c }
\hat 1_6 &  -\Delta \hat Z \\
0 & \hat 1_2
\end{array}\right),
\ee
where $\hat 1_{6,2}$ are $6\times 6$ and $2\times 2$ identity matrices and
$\Delta\hat Z$ is a $6\times 2$ matrix. Imposing the condition that the
renormalized operators coincide in the two schemes, $\Delta\hat Z$
is defined as the difference of
the vectors $\vec Z$ of eq. (\ref{cz}) calculated in the two different
regularization schemes, i.e. $ \Delta \hat Z= (\vec Z_7^a-\vec Z_7^b,
\vec Z_8^a-\vec Z_8^b)$.

At the order we are interested in, eq. (\ref{check}) gives
\be
\Delta\hat\gamma=\hat\gamma^a-\hat\gamma^b=
[\Delta\hat Z,\hat\gamma^b]-\Delta\hat Z\hat\gamma^b\Delta\hat Z.
\label{check1}
\ee
Contrary to the already mentioned $\Delta S=1$ case, eq. (\ref{check1})
does not define a scheme independent combination of the one and two
loop matrices, namely it can not be written just in terms of quantities which
are either calculated as differences between the two schemes or scheme
independent by themselves. However a scheme independent combination actually
exists due to the structure of the matrices. In fact, applying eq.
(\ref{check1}) to anomalous dimension matrices of the form of eq.
(\ref{matform}), we find
\bea
\hat\gamma_r^a&=&\hat\gamma^b_r=\hat\gamma_r\nn\\
(\Delta \vec \beta_7)_j=
\Delta\gamma_{j7}&=&\left[\left(\gamma_{77}\hat 1-\hat\gamma_r\right)\Delta\vec
Z_7+\gamma_{87}\Delta\vec Z_8 \right]_j\nn \\
(\Delta \vec \beta_8)_j=
\Delta\gamma_{j8}&=&\left[\left(\gamma_{88}\hat 1-\hat\gamma_r\right)
\Delta\vec Z_8\right]_j,
\label{check2}
\eea
where $j=1,\dots,6$. This clearly implies that the combinations
\bea
&&\vec\beta_7-\left(\gamma_{77}\hat 1-\hat\gamma_r\right)\vec Z_7-
\gamma_{87}\vec Z_8 \nn \\
&&\vec \beta_8-\left(\gamma_{88}\hat 1-\hat\gamma_r\right)\vec Z_8
\eea
are scheme independent.
Using eqs. (\ref{check2}), one can easily check that the combinations
$\vec\omega_7,\, \vec\omega_8$ in eq. (\ref{omega}) give $\Delta
\vec\omega_{7,8}=0$, i.e. they are scheme independent too. In turn this implies
that, as expected, the Effective Hamiltonian, eq. (\ref{heff}), is independent
of the regularization scheme.

Eqs. (\ref{check2}) can be used to relate classes of diagrams calculated in
different schemes. Thus it can be a useful check of the calculation, see
Sec. \ref{sec:diag} below.

\section {Dimensional Regularization Schemes}
\label{sec:schemes}

In this section we briefly recall the definitions of the regularization
procedures we have used for our calculations. Other nice discussions can be
found in refs. \cite{bw1,acmp} and references therein, while for more details
the reader should refer to the literature on the specific subject.

The regularization procedures we have used, NDR, HV and DRED, are all based on
the dimensional regularization of the loop integrals \cite{hv}, which
consists in performing the integration over the loop momenta in $D$ dimensions,
thus turning the divergences into regularized $1/(4-D)$ terms. The extension of
the Dirac algebra to non-integer dimensions presents no difficulty except
for the properties of traces involving $\gamma_5$ and in general
expressions containing the completely antisymmetric tensor
$\epsilon_{\mu\nu\rho\sigma}$,
which does not have any meaningful extension in $D$ dimensions. Actually
the three schemes differ from each other in the way they treat $\gamma_5$.
Another relevant point is that the relation
\be
\gamma_\mu\gamma_\nu\gamma_\rho=\gamma_\mu g_{\nu\rho}-\gamma_\nu g_{\mu\rho}
+\gamma_\rho g_{\mu\nu}-i\gamma^\sigma\gamma_5\epsilon_{\mu\nu\rho\sigma}
\ee
which, in 4 dimensions, projects the product of three $\gamma$ matrices on
the basis does not hold in $D$ dimensions. This implies that more complicated
tensor structures, which eventually vanish in 4 dimensions, appear in the
regularized theory and must be considered in the renormalization procedure.

Let us now separately discuss how the three schemes are implemented.

\subsection {Naive Dimensional Regularization}
\label{sec:ndr}
In NDR all the Lorentz indices appearing in the regularized theory are
assumed to be in $D$ dimensions. The Dirac algebra is identical to the
4-dimensional one, including the properties of $\gamma_5$,
once the 4-dimensional metric tensor is replaced by the $D$-dimensional one.
The definitions we are interested in are simply
\bea
g_{\mu\nu}g^{\mu\nu}&=&D\nn\\
\left\{\gamma_\mu,\gamma_5\right\}&=&0.
\eea
Obviously, since the completely antisymmetric tensor is not defined in
$D$-dimensions, traces involving odd number of $\gamma_5$ are ill-defined too.
However, to our knowledge, once the problem is properly fixed, there is no
calculation which shows a failure of this scheme.
Our way to fix the $\gamma_5$ problem is the one implemented by Schoonschip
\cite{schip},
which defines traces of odd number of $\gamma_5$ in NDR in the same way they
are
defined in the HV scheme, see below.

\subsection {'t Hooft-Veltman Regularization}
\label{sec:hv}
In HV \cite{hv}, Lorentz indices in $D$, 4 and $(D-4)$ dimensions are
introduced,
together with the corresponding metric tensors $g_{\mu\nu}$, $\tilde
g_{\mu\nu}$
and $\hat g_{\mu\nu}$. All the $\gamma$ matrices are taken in $D$ dimension,
then
indices are split in 4 and $(D-4)$ components, according to the rules
\bea
g_{\mu\nu}&=&\tilde g_{\mu\nu}+\hat g_{\mu\nu}\nn\\
\tilde g_{\mu\nu}\tilde g^{\mu\nu}=4&,&\hat g_{\mu\nu}\hat g^{\mu\nu}=D-4\nn\\
\tilde g_{\mu\nu}\hat g^{\nu\rho}&=&0.
\eea
These rules define also the extended Dirac algebra, once one notes that
$\gamma$
matrices in 4 ($\tilde\gamma^\mu$) and $(D-4)$ ($\hat\gamma^\mu$) dimensions
can be written in terms of $D$ dimensional matrices as $\tilde g^{\mu\nu}
\gamma_\nu$ and $\hat g^{\mu\nu}\gamma_\nu$ respectively and the usual
commutation relations among the $D$-dimensional Dirac matrices in terms of the
$D$-dimensional metric tensor $g^{\mu\nu}$ are assumed.

$\gamma$ matrices in $D$ dimensions do not have definite commutation
relation with $\gamma_5$. In fact the following relations hold in HV
\be
\left\{\tilde\gamma_\mu,\gamma_5\right\}=0\qquad,
\qquad\left[\hat\gamma_\mu,\gamma_5\right]=0.
\ee
This is equivalent to define $\gamma_5$ as the product $i\tilde\gamma_0
\tilde\gamma_1\tilde\gamma_2\tilde\gamma_3$. This way of treating
$\gamma_5$ in $D$ dimensions is the only
known one which does not give rise to algebraic inconsistencies or
introduce ill-defined quantities. On the other hand, due to the splitting
of indices, the HV scheme is the most difficult to handle among the three
considered here as far as the algebraic manipulation problems are concerned.

Finally we mention that the chiral vertices in $D$ dimensions can be
defined in different ways, all having the same limit when $D$ tends to 4
dimensions. We use the symmetrized form
\be
\frac{1}{2}(V\pm A)\gamma_\mu (V\pm A)=\tilde\gamma_\mu (V\pm A).
\ee
In this way the bare vertices preserve the chirality of the external
fields also in $D$ dimensions.

\subsection {Dimensional Reduction}
\label{sec:dred}
In DRED \cite{sieg} the Dirac matrices have indices in 4 dimensions
only, thus the algebra is greatly simplified.
The $D$-dimensional indices are introduced by the loop integrals, which
generate the $D$-dimensional metric tensor $g_{\mu\nu}$, so that $\gamma_\mu$
in $D$ dimensions is given by $g_\mu^\nu\tilde\gamma_\nu$. The basic
rules
\bea
\tilde g_{\mu\nu}&=& g_{\mu\nu}+\hat g_{\mu\nu}\nn\\
g_{\mu\nu} g^{\mu\nu}=D&,&\hat g_{\mu\nu}\hat g^{\mu\nu}=4-D\nn\\
g_{\mu\nu}\hat g^{\nu\rho}&=&0
\eea
are formally similar to the HV ones provided the roles of $g_{\mu\nu}$ and
$\tilde g_{\mu\nu}$ are exchanged, but, contrary to that case, DRED is
known to be algebraically inconsistent \cite{maison}.
Moreover one should also mention that DRED fails to reproduce the triangle
anomaly, unless further ad-hoc prescriptions are assumed \cite{nictow}. Finally
it is well known that one has to take care of many theoretical
subtleties regarding the renormalization of operators, living in
$(4-D)$ dimensions, when higher order calculations are performed
\cite{acmp}.

%___________________________________________________________________________
\begin{figure}[t]   % produce figure here
    \begin{center}
       \setlength{\unitlength}{1truecm}
       \begin{picture}(6.0,9.0)
          \put(-4.2,-2.5){\special{1loop_op.ps}}
       \end{picture}
    \end{center}
\caption[]{One loop diagrams which generate both
the 4-fermion operator mixing and the counter-terms to be used in
the two loop calculation.}
\protect\label{fig:1loop_op}
\end{figure}
%___________________________________________________________________________

In spite of all these problems, no calculation, with the exception of the
triangle anomaly, gives so far a wrong result. Concerning our calculation,
neither inconsistencies of the regularization scheme nor other problems with
the
renormalization procedure in $D$ dimensions show out, once a suitable
definition of the $(4-D)$-dimensional operators is introduced, as explained
in Sec. \ref{sec:diag}.

\section {Diagrams and Counter-terms}
\label{sec:diag}
In this section the main results of our study are given. Contributions from
the one and two loop Feynman diagrams in figs. (\ref{fig:1loop_op})-
(\ref{fig:2loop}) to the anomalous dimension matrix are presented in the three
considered regularization schemes. Tabs.
(\ref{tab:1loop_op})-(\ref{tab:2loop_frr}) contain the results of these
diagrams in terms of the coefficients of the poles in the number of dimensions,
appearing as powers of $1/\epsilon=2/(4-D)$.

\subsection {One Loop Diagrams}
\label{sec:onediag}

%---------------------------------------------------------------
\begin{table}
\begin{center}
\begin{tabular}{|c|c|c|c|}\hline\hline
\multicolumn{1}{|c}{diagram} & \multicolumn{1}{|c}{M}
&\multicolumn{1}{|c|}{$\LL$} &
\multicolumn{1}{c|}{$\LR$}
\\
\hline
 & & $(1/\epsilon)$ & \phantom{m}$(1/\epsilon)$\\\hline
$V_1$ & 2&1 &1\\
$V_2$ &2& -4& -1  \\
$V_3$ &2& 1 & 4  \\
$V_4$ & 1&-4/3 & - \\
$V_5$ & 1&-4/3 & -4/3 \\
\hline\hline
\end{tabular}
\caption[]{Singular parts of the diagrams in
figs. (\ref{fig:1loop_op}), with a $\LL$ or a $\LR$ vertex insertion.
$V_1$-$V_3$ results are proportional to the inserted 4-fermion structures.
$V_4$-$V_5$ are proportional to $\LV$.
The multiplicity of the diagrams is also reported in the table.
Colour factors and $\alphas / 4 \pi$ are omitted.}
\label{tab:1loop_op}
\end{center}
\end{table}
%---------------------------------------------------------------

Let us start with the $O(\alphas^0)$ finite mixing among 4-fermion and
magnetic operators, already discussed in Sec.
\ref{sec:solut}. The penguin diagrams in fig. (\ref{fig:1loop_ct})
could mix 4-fermion operator with magnetic ones.
The coupling constant is re-absorbed in the definition of the magnetic
operators, so this mixing appears already at $O(\alphas^0)$.
However penguin diagrams in HV and DRED have
neither pole nor finite parts on the magnetic form factor. Differently in NDR,
when massive quark fields propagate in the loop, the diagram $P1$ induces a
finite
mixing among $Q_5,\,Q_6$ and $Q_7,\, Q_8$. In fact the calculation of
$P1$ with a $\LR$ vertex insertion gives
\be
2 m_b \left(\qslash\gamma_\mu
-\gamma_\mu\qslash)\right)\left(1+\gamma_5\right).
\ee
Selecting the magnetic form factor through the projection
\be
m_b (1+\gamma_5)\qslash \tilde\gamma_\mu \to \frac{1}{2},\qquad
m_b (1+\gamma_5)\tilde\gamma_\mu \qslash \to -\frac{1}{2},
\label{prj1}
\ee
the values of the vector $\vec Z$ in eq. (\ref{zv}) can be readily obtained.
The
accidental vanishing of $\vec Z$ in both the HV and DRED schemes makes them to
give coincident results even at intermediate stages of the calculation.

%___________________________________________________________________________
\begin{figure}[t]   % produce figure here
    \begin{center}
       \setlength{\unitlength}{1truecm}
       \begin{picture}(6.0,9.0)
          \put(-4.2,-1){\special{1loop_mag.ps}}
       \end{picture}
    \end{center}
    \caption[]{Diagrams responsible for the one loop renormalization of the
magnetic operators.}
\protect\label{fig:1loop_mag}
\end{figure}
%___________________________________________________________________________

Figs. (\ref{fig:1loop_op}) and (\ref{fig:1loop_mag}) show all the diagrams
required to calculate at the leading order
the 4-fermion operator mixing matrix $\hat\gamma_r$ and the mixing of the
magnetic operators among themselves. The corresponding results can be found
in tabs. (\ref{tab:1loop_op}) and (\ref{tab:1loop_mag}), where the pole
coefficients are reported. These results, which do not depend on the
regularization scheme, are established since a long time
\cite{svz,gw}, thus we omit
the details on their calculation. Our results, as well as the
corresponding
anomalous dimension sub-matrices in eqs. (\ref{gamma_r}) and (\ref{gamma_mag}),
coincide with those of refs. \cite{gw} and
\cite{svz}. We just mention that the gauge dependence of the diagrams
$M_1$-$M_5$ cancels out when the
external gluon field renormalization, see eqs. (\ref{gammao})-(\ref{gamext}),
is taken into account.

%---------------------------------------------------------------
\begin{table}
\begin{center}
\begin{tabular}{|c|c|} \hline \hline
$Diagram$ & $1/\epsilon$ \\ \hline
\tsp$M_{\scriptscriptstyle 1}$ & $-$\\
\tsp$M_{\scriptscriptstyle 2}$ & -4 \\
\tsp$M_{\scriptscriptstyle 3}$ & -1 \\
\tsp$M_{\scriptscriptstyle 4}$ & 6 \\
\tsp$M_{\scriptscriptstyle 4}^{\scriptscriptstyle bg}$ & 5 \\
\tsp$M_{\scriptscriptstyle 5}$ & $-\frac{3}{2}$ \\
\tsp$M_{\scriptscriptstyle 5}^{\scriptscriptstyle bg}$ & -2 \\
\hline \hline
\end{tabular}
\caption[]{Singular parts of the diagrams in fig. (\ref{fig:1loop_mag}).
All the results are proportional to the magnetic operators. The factor
$\alphas / 4\pi$ is omitted. The diagrams
$M_4$ and $M_5$ are calculated both in the Feynman and background gauges.}
\label{tab:1loop_mag}
\end{center}
\end{table}
%---------------------------------------------------------------

\subsection {One Loop Counter-terms}
\label{sec:onecntr}

Counter-terms to be inserted in the two loop diagrams are obtained from the
diagrams in figs. (\ref{fig:1loop_op})-(\ref{fig:1loop_ct}).
Many scheme dependent operators, vanishing in 4 dimensions, are generated
along with the usual counter-terms proportional to the
operators already present in the 4-dimensional basis.
It is well known \cite{cfmr2,bw1,acmp}
that ``effervescent'' operators must be considered in the renormalization
procedure, namely they must be properly inserted as counter-terms in the two
loop diagrams, in order to have the right final result.

%___________________________________________________________________________
\begin{figure}[t]   % produce figure here
    \begin{center}
       \setlength{\unitlength}{1truecm}
       \begin{picture}(6.0,9.0)
          \put(-6.2,-1.5){\special{1loop_ct.ps}}
       \end{picture}
    \end{center}
\caption[]{Diagrams which generate the counter-terms discussed
in the text. In DRED they generate ``effervescent''
contributions that cannot be omitted.}
\protect\label{fig:1loop_ct}
\end{figure}
%___________________________________________________________________________

Here we do not repeat the procedure to define these operators in HV and NDR,
since it is fully explained in refs. \cite{cfmr2,bjlw1,bw1}. Also the DRED
``effervescent''
operators generated by the 4-fermion diagrams in fig. (\ref{fig:1loop_op})
are known from ref. \cite{acmp}. We confirm all previous results and give some
details only on our DRED calculation of the ``effervescent'' operators due
to the penguin and gluon-photon diagrams in fig. (\ref{fig:1loop_ct}), recently
presented also in ref. \cite{mis3}.

Contrary to other schemes, in DRED the $\gamma$ algebra is performed in
4 dimensions so that no complicated tensor structure appears and the
``effervescent'' operators can be readily defined by inspection as
those terms that have $(4-D)$-dimensional Lorentz indices saturated on
the external fields.
The ``effervescent'' part of the penguin diagrams in DRED is then given by
\be
-\frac{2}{3}\ep q^2\hat g^{\mu\nu}\gamma_\nu\left(1-\gamma_5\right).
\ee
This result coincides with the corresponding one of ref. \cite{mis3}.

Concerning the diagrams $C_{1a}$ and $C_{1b}$ in fig. (\ref{fig:1loop_ct}),
their ``effervescent'' parts are found to be respectively
\bea
&& -\frac{2}{3\epsilon} (q+2l)^\mu \gamma_\nu \hat g^{\nu\rho}
(1-\gamma_5)+\ep \left(\qslash\gamma^\mu-\gamma^\mu\qslash\right)
\gamma_\nu\hat g^{\nu\rho}(1-\gamma_5)+\dots\nn\\
&& \frac{2}{3\epsilon} (q+2l)^\mu \gamma_\nu \hat g^{\nu\rho}
(1-\gamma_5)+\ep \left(\qslash\gamma^\mu-\gamma^\mu\qslash\right)
\gamma_\nu\hat g^{\nu\rho}(1-\gamma_5)+\dots
\label{2gluons}
\eea
The dots indicates further terms that can be
omitted since they contribute only to the ``effervescent'' part of the
two loop diagrams.
By summing the two terms in eq. (\ref{2gluons}), we obtain a
counter-term with a pole coefficient which disagrees from the one presented
in ref. \cite{mis3}.
We believe that there is a misprint in ref. \cite{mis3}, since we
obtain the same results reported there for the insertions of this
``effervescent'' operator in the two loop diagrams.

It is worthwhile to note that, as shown in ref.
\cite{cfmr2}, the insertions of the counter-terms generated by the diagrams
in fig. (\ref{fig:1loop_ct}) and by the longitudinal parts of the penguin
diagrams vanish, so that one can retain 4-fermion operators only as
counter-terms in the Hamiltonian. However in DRED the diagrams of fig.
(\ref{fig:1loop_ct}) give ``effervescent'' contributions that must be
taken into account.

\subsection {Two Loop Diagrams}
\label{sec:twodia}

The relevant two loop diagrams are shown in fig.
(\ref{fig:2loop}) and the corresponding results in the three considered
regularization schemes can be found in tabs.
(\ref{tab:2loop_plb})-(\ref{tab:2loop_frr}), which contain the pole
coefficients of the diagrams projected on the magnetic form factor.
Two tables for each different
kind of Dirac structure inserted in the upper vertex are presented. The
first one contains the pole coefficients of the bare diagram ($D$),
the insertion of 4-dimensional counter-terms ($C$) and the insertion of
``effervescent'' counter-terms ($E$). The second one collects the final results
of the renormalized diagrams ($\overline{D}$), obtained as
\be
\overline{D}=D-C-\frac{1}{2}E,
\label{rendiag}
\ee
see eq. (\ref{expgam}) below.
Results from the insertion of a $\LL$ upper vertex in the $P$-type diagrams
of fig. (\ref{fig:2loop}) are reported in tabs.
(\ref{tab:2loop_plb})-(\ref{tab:2loop_plr}).
Those coming from $\LL$ insertion in $F$-type diagrams can be found in tabs.
(\ref{tab:2loop_flb})-(\ref{tab:2loop_flr}).
$\LR$ 4-fermion vertex, inserted in $P$- and $F$-type diagrams, originate the
results in tabs. (\ref{tab:2loop_prb})-(\ref{tab:2loop_prr}) and
(\ref{tab:2loop_frb})-(\ref{tab:2loop_frr}) respectively.
Note that the left-right operator insertion in the $P$-type diagrams, tab.
(\ref{tab:2loop_prb}), do not vanish just because of the massive loop
propagators,
thus they contribute only when $q=b$ is taken in $Q_5,\, Q_6$.

%___________________________________________________________________________
\begin{figure}[t]   % produce figure here
    \begin{center}
       \setlength{\unitlength}{1truecm}
       \begin{picture}(7.0,16.8)
          \put(-5.2,-1){\special{2loop.ps}}
       \end{picture}
    \end{center}
\caption[]{The two loop diagrams relevant for the calculation of the
vectors $\vec \beta_{7,8}$ in eq. (\ref{matform}).}
\protect\label{fig:2loop}
\end{figure}
%___________________________________________________________________________

The calculation of the two loop diagrams result in many tensor structures,
containing $m_b$ mass and/or external momenta. Further projections are
required, besides those in eq. (\ref{prj1})
\bea
m_b(1+\gamma_5)\tilde\gamma^\mu\pslash \to 0&,&
m_b(1+\gamma_5)\pslash \tilde\gamma^\mu \to \frac{1}{2}\nn\\
m_b(1+\gamma_5)\frac{\pslash\qslash}{q^2}q^\mu \to 0&,&
m_b(1+\gamma_5)\frac{\pslash\qslash}{q^2}p^\mu \to \frac{1}{4}\nn\\
m_b(1+\gamma_5)\frac{\qslash\pslash}{q^2}q^\mu \to 0&,&
m_b(1+\gamma_5)\frac{\qslash\pslash}{q^2}p^\mu \to 0\nn\\
m_b(1+\gamma_5)p^\mu \to \frac{1}{4}&,&
m_b(1+\gamma_5) q^\mu \to 0\nn\\
\nn\\
(1+\gamma_5)\qslash q^\mu \to 0&,&
(1+\gamma_5)\qslash p^\mu \to -\frac{1}{4}\nn\\
(1+\gamma_5)\pslash q^\mu \to 0&,&
(1+\gamma_5)\pslash p^\mu \to -\frac{1}{4}.
\label{prj2}
\eea
Moreover in DRED also the antisymmetric tensor $\epsilon_{\mu\nu\rho\sigma}$
appears, either in pure 4-dimensional expressions or in tensors involving
$\hat g_{\mu\nu}$. In these cases the projections are
\bea
m_b (1+\gamma_5)\tilde\gamma_\nu\tilde\gamma_\rho
\epsilon^{\mu\nu\rho\sigma}q_\sigma &\to& 1\nn\\
m_b (1+\gamma_5)\tilde\gamma_\nu\tilde\gamma_\rho
\epsilon^{\mu\nu\rho\sigma}p_\sigma &\to& -1\nn\\
m_b (1+\gamma_5) \tilde\gamma_\alpha\tilde\gamma_\rho\hat g^{\alpha}_{\nu}
\epsilon^{\mu\nu\rho\sigma}q_\sigma&\to& -\frac{1}{2}(4-D)\nn\\
m_b (1+\gamma_5)\tilde\gamma_\alpha\tilde\gamma_\rho\hat g^{\alpha}_{\nu}
\epsilon^{\mu\nu\rho\sigma}p_\sigma&\to& -\frac{1}{4}(4-D)\nn\\
\nn\\
(1+\gamma_5)\tilde\gamma_\nu\epsilon^{\mu\nu\rho\sigma}
q_\rho p_\sigma &\to& \frac{1}{4}\nn\\
(1+\gamma_5)\tilde\gamma_\alpha\hat g^\alpha_\nu
\epsilon^{\mu\nu\rho\sigma}q_\rho p_\sigma &\to& -\frac{1}{4}(4-D).
\label{prj_dred}
\eea

Let us consider now the relations among different schemes. Eqs. (\ref{check2})
enforce a set of relations among one and two loop diagrams, which can be
easily obtained by considering how each diagram contributes to the anomalous
dimension matrix (i.e. its Dirac and colour structure). These relations are
\bea
&&\Delta (P_2+P_3) = 0,\qquad
\Delta (F_2+F_3) = 0\nn\\
&&\Delta (P_2+P_4) = 0,\qquad
\Delta (F_2+F_4) = 0\nn\\
&&\Delta P_7 = -\frac{1}{4} P_1 \Delta P^{m_b}_1,\qquad
\Delta F_7 = -\frac{1}{4} F_1 \Delta P^{m_b}_1\nn\\
&&\Delta (P_2^{m_b}+P_3^{m_b}+P_5^{m_b}+P_7^{m_b}+P_8^{m_b}+P_9^{m_b}) =
-7\Delta P_1^{m_b}\nn\\
&&\Delta (P_5^{m_b}+P_8^{m_b}+P_9^{m_b}) = -5\Delta P_1^{m_b}\nn\\
&&\Delta (P_2^{m_b}+P_4^{m_b}+P_6^{m_b}+P_8^{m_b}) = -2\Delta P_1^{m_b},
\label{check3}
\eea
where $\Delta$ indicates the difference between two different regularization
schemes. Quantities denoted by $m_b$ refer to the results of the left-right
$P$-type diagrams with a mass insertion into the loop propagators.
They can be found in tab. (\ref{tab:2loop_prr}), with the exception of
$P_1^{m_b}$, which is equal to 2 in NDR and vanishes in the other two schemes.
As already noted, HV and DRED calculations coincide diagram by diagram,
so that eqs. (\ref{check3}) are not really interesting. However, when comparing
HV or DRED with NDR, this check is effective and one can verify that all our
diagrams satisfy eqs. (\ref{check3}).

Among the other checks passed by our results, one can readily verify
\begin{itemize}
\item the cancellation of all the double poles, as an indication that this
is a leading order calculation, see eqs. (\ref{expgam});
\item the usual relation between the double poles of the bare diagram and
the insertion of the counter-terms, the second being two times larger then
the first;
\item the gauge independence, checked by using the Feynman and the
background gauges \cite{abbott}.
\end{itemize}

\newpage
\clearpage

\begin{table}
\begin{center}
\begin{tabular}{|c|cccc|cccc|ccc|}
\hline\hline
\multicolumn{1}{|c} { } &
\multicolumn{4}{|c}{$D$} &
\multicolumn{4}{|c}{$C$} &
\multicolumn{3}{|c|}{$E$} \\
\hline
\multicolumn{1}{|c|}{$T_N$}&
$\frac{1}{\epdue}$ & $\tsq{\frac{1}{\epuno}}^{\sss HV}$ &
${\frac{1}{\epuno}}^{\sss NDR}$ &  ${\frac{1}{\epuno}}^{\sss DRED}$ &
$\frac{1}{\epdue}$ &
${\frac{1}{\epuno}}^{\sss HV}$ & ${\frac{1}{\epuno}}^{\sss NDR}$ &
 ${\frac{1}{\epuno}}^{\sss DRED}$ &
${\frac{1}{\epuno}}^{\sss HV}$ & ${\frac{1}{\epuno}}^{\sss NDR}$ &
${\frac{1}{\epuno}}^{\sss DRED}$ \\[0.5ex]
\hline
\tsp$P_2$ &
-$\frac{1}{9}$ & -$\frac{91}{54}$ & -$\frac{97}{54}$ & -$\frac{29}{27}$ &
-$\frac{2}{9}$ & -$\frac{10}{27}$ & -$\frac{22}{27}$ & -$\frac{10}{27}$ &
$-$ & $-$ & $\frac{11}{9}$ \\
\tsp$P_3$ &
$\frac{1}{9}$ & -$\frac{71}{54}$ & -$\frac{65}{54}$ & $-\frac{25}{27}$ &
$\frac{2}{9}$ &  $\frac{10}{27}$ & $\frac{22}{27}$ & $\frac{10}{27}$ &
$-$ & $-$ & $\frac{7}{9}$ \\
\tsp$P_4$ &
$\frac{1}{9}$ & $\frac{25}{54}$ & $\frac{31}{54}$ & $\frac{19}{54}$ &
$\frac{2}{9}$ & $\frac{10}{27}$ & $\frac{22}{27}$ & $\frac{10}{27}$ &
$-$ & $-$ & -$\frac{2}{9}$ \\
\tsp$P_{4}^{bg}$ &
$\frac{1}{9}$ & $\frac{25}{54}$ & $\frac{31}{54}$ & $\frac{19}{54}$ &
$\frac{2}{9}$ & $\frac{10}{27}$ & $\frac{22}{27}$ & $\frac{10}{27}$ &
$-$ & $-$ & -$\frac{2}{9}$ \\
\tsp$P_7$ &
$-$ & -$\frac{2}{9}$ & -$\frac{8}{9}$ & $\frac{1}{9}$ &
$-$ & $-$ & -$\frac{4}{3}$ & $-$ &
$-$ & $-$ & $\frac{2}{3}$ \\[0.5ex]
\hline\hline
\end{tabular}
\caption[]{Singular parts of the $P$ diagrams in fig. (\ref{fig:2loop}) with
a $\LL$ vertex insertion. The common double poles and the single poles,
calculated in HV, NDR and DRED,
are presented for the bare diagrams ($D$), the 4-dimensional
($C$) and the ``effervescent'' ($E$) counter-terms. All the results are
proportional to the magnetic operators. Diagram $P_4$ is calculated both
in the Feynman and background gauges.}
\label{tab:2loop_plb}
\end{center}
\end{table}

\begin{table}
\begin{center}
\begin{tabular}{|c|cccc|}
\hline\hline
\multicolumn{1}{|c} { } &
\multicolumn{4}{|c|}{\tsq$\overline{D}$} \\
\hline
\multicolumn{1}{|c|}{$T_N$}&
$\frac{1}{\epdue}$ & $\tsq{\frac{1}{\epuno}}^{\sss HV}$ &
${\frac{1}{\epuno}}^{\sss NDR}$ & ${\frac{1}{\epuno}}^{\sss DRED}$ \\[0.5ex]
\hline
\tsp$P_2$ &
$\frac{1}{9}$ & -$\frac{71}{54}$ & -$\frac{53}{54}$ & -$\frac{71}{54}$ \\
\tsp$P_3$ &
-$\frac{1}{9}$ & -$\frac{91}{54}$ & -$\frac{109}{54}$ & -$\frac{91}{54}$ \\
\tsp$P_4$ &
-$\frac{1}{9}$ & $\frac{5}{54}$ & -$\frac{13}{54}$ & $\frac{5}{54}$ \\
\tsp$P_{4}^{bg}$ &
-$\frac{1}{9}$ & $\frac{5}{54}$ & -$\frac{13}{54}$ & $\frac{5}{54}$ \\
\tsp$P_7$ &
$-$ & -$\frac{2}{9}$ & $\frac{4}{9}$ & -$\frac{2}{9}$ \\[0.5ex]
\hline\hline
\end{tabular}
\caption[]{Singular parts of the renormalized $P$ diagrams in fig.
(\ref{fig:2loop}) with a $\LL$ vertex insertion. These results are obtained
from tab. (\ref{tab:2loop_plb}) using eq. (\ref{rendiag}).}
\label{tab:2loop_plr}
\end{center}
\end{table}

\newpage
\clearpage

\begin{table}
\begin{center}
\begin{tabular}{|c|cccc|cccc|ccc|}
\hline\hline
\multicolumn{1}{|c} { } &
\multicolumn{4}{|c}{$D$} &
\multicolumn{4}{|c}{$C$} &
\multicolumn{3}{|c|}{$E$} \\
\hline
\multicolumn{1}{|c|}{$T_N$}&
$\frac{1}{\epdue}$ & $\tsq{\frac{1}{\epuno}}^{\sss HV}$ &
${\frac{1}{\epuno}}^{\sss NDR}$ &  ${\frac{1}{\epuno}}^{\sss DRED}$ &
$\frac{1}{\epdue}$ &
${\frac{1}{\epuno}}^{\sss HV}$ & ${\frac{1}{\epuno}}^{\sss NDR}$ &
 ${\frac{1}{\epuno}}^{\sss DRED}$ &
${\frac{1}{\epuno}}^{\sss HV}$ & ${\frac{1}{\epuno}}^{\sss NDR}$ &
${\frac{1}{\epuno}}^{\sss DRED}$ \\[0.5ex]
\hline
\tsp$F_2$ &
-$\frac{1}{9}$ & -$\frac{91}{54}$ & -$\frac{103}{54}$ & -$\frac{29}{27}$ &
-$\frac{2}{9}$ & -$\frac{10}{27}$ & -$\frac{22}{27}$ & -$\frac{10}{27}$ &
$-$ & $-$ & $\frac{11}{9}$ \\
\tsp$F_3$ &
$\frac{1}{9}$ & -$\frac{71}{54}$ & -$\frac{59}{54}$ & $-\frac{25}{27}$ &
$\frac{2}{9}$ &  $\frac{10}{27}$ & $\frac{22}{27}$ & $\frac{10}{27}$ &
$-$ & $-$ & $\frac{7}{9}$ \\
\tsp$F_4$ &
$\frac{1}{9}$ & $\frac{25}{54}$ & $\frac{37}{54}$ & $\frac{19}{54}$ &
$\frac{2}{9}$ & $\frac{10}{27}$ & $\frac{22}{27}$ & $\frac{10}{27}$ &
$-$ & $-$ & -$\frac{2}{9}$ \\
\tsp$F_{4}^{bg}$ &
$\frac{1}{9}$ & $\frac{25}{54}$ & $\frac{37}{54}$ & $\frac{19}{54}$ &
$\frac{2}{9}$ & $\frac{10}{27}$ & $\frac{22}{27}$ & $\frac{10}{27}$ &
$-$ & $-$ & -$\frac{2}{9}$ \\
\tsp$F_7$ &
$-$ & -$\frac{2}{9}$ & -$\frac{8}{9}$ & $\frac{1}{9}$ &
$-$ & $-$ & -$\frac{4}{3}$ & $-$ &
$-$ & $-$ & $\frac{2}{3}$ \\[0.5ex]
\hline\hline
\end{tabular}
\caption[]{The same as tab. (\ref{tab:2loop_plb}) for the $F$ diagrams
with a $\LL$ vertex insertion.}
\label{tab:2loop_flb}
\end{center}
\end{table}

\begin{table}
\begin{center}
\begin{tabular}{|c|cccc|}
\hline\hline
\multicolumn{1}{|c} { } &
\multicolumn{4}{|c|}{\tsq$\overline{D}$} \\
\hline
\multicolumn{1}{|c|}{$T_N$}&
$\frac{1}{\epdue}$ & $\tsq{\frac{1}{\epuno}}^{\sss HV}$ &
${\frac{1}{\epuno}}^{\sss NDR}$ & ${\frac{1}{\epuno}}^{\sss DRED}$ \\[0.5ex]
\hline
\tsp$F_2$ &
$\frac{1}{9}$ & -$\frac{71}{54}$ & -$\frac{59}{54}$ & -$\frac{71}{54}$ \\
\tsp$F_3$ &
-$\frac{1}{9}$ & -$\frac{91}{54}$ & -$\frac{103}{54}$ & -$\frac{91}{54}$ \\
\tsp$F_4$ &
-$\frac{1}{9}$ & $\frac{5}{54}$ & -$\frac{7}{54}$ & $\frac{5}{54}$ \\
\tsp$F_{4}^{bg}$ &
-$\frac{1}{9}$ & $\frac{5}{54}$ & -$\frac{7}{54}$ & $\frac{5}{54}$ \\
\tsp$F_7$ &
$-$ & -$\frac{2}{9}$ & $\frac{4}{9}$ & -$\frac{2}{9}$ \\[0.5ex]
\hline\hline
\end{tabular}
\caption[]{The same as tab. (\ref{tab:2loop_plr}) for the $F$ diagrams
with a $\LL$ vertex insertion.}
\label{tab:2loop_flr}
\end{center}
\end{table}

\newpage
\clearpage

\begin{table}
\begin{center}
\begin{tabular}{|c|cccc|cccc|ccc|}
\hline\hline
\multicolumn{1}{|c} { } &
\multicolumn{4}{|c}{$D$} &
\multicolumn{4}{|c}{$C$} &
\multicolumn{3}{|c|}{$E$} \\
\hline
\multicolumn{1}{|c|}{$T_N$}&
$\frac{1}{\epdue}$ & $\tsq{\frac{1}{\epuno}}^{\sss HV}$ &
${\frac{1}{\epuno}}^{\sss NDR}$ &  ${\frac{1}{\epuno}}^{\sss DRED}$ &
$\frac{1}{\epdue}$ &
${\frac{1}{\epuno}}^{\sss HV}$ & ${\frac{1}{\epuno}}^{\sss NDR}$ &
 ${\frac{1}{\epuno}}^{\sss DRED}$ &
${\frac{1}{\epuno}}^{\sss HV}$ & ${\frac{1}{\epuno}}^{\sss NDR}$ &
${\frac{1}{\epuno}}^{\sss DRED}$ \\[0.5ex]
\hline
\tsp$P_2$ &
$-$ & $2$ & $3$ & $2$ &
 $-$ & $-$ & $4$ & $-$ &
 $-$ & $-$ & $-$ \\
\tsp$P_3$ &
 $-$ & $2$ & -$7$ & $2$ &
 $-$ & $-$ & -$20$ & $-$ &
 $-$ & $16$ & $-$ \\
\tsp$P_4$ &
 $-$ & $-$ & $-6$ & $-$ &
 $-$ & $-$ & $-$ & $-$ &
 $-$ & $-$ & $-$ \\
\tsp$P_{4}^{bg}$ &
 $-$ & $-$ & $-5$ &  $-$ &
 $-$ & $-$ & $-$ & $-$ &
 $-$ & $-$ & $-$ \\
\tsp$P_5$ &
 $-$ & $-$ & $-$ &  $-$ &
 $-$ & $-$ & $-$ & $-$ &
 $-$ & $8$ & $-$ \\
\tsp$P_6$ &
 $-$ & $-$ & $-$ &  $-$ &
 $-$ & $-$ & $-6$ & $-$ &
 $-$ & $-$ & $-$ \\
\tsp$P_{6}^{bg}$ &
 $-$ & $-$ & $3$ &  $-$ &
 $-$ & $-$ & $-2$ & $-$ &
 $-$ & $-$ & $-$ \\
\tsp$P_7$ &
 $-$ & -$\frac{2}{3}$ & $-4$ & $-$ &
 $-$ & $-$ & $-$ & $-$ &
 -$\frac{4}{3}$ & $-$ & $-$ \\
\tsp$P_8$ &
 $-$ & $-$ & $1$ &  $-$ &
 $-$ & $-$ & $2$ & $-$ &
 $-$ & $-$ & $-$ \\
\tsp$P_9$ &
 $-$ & $-$ & $1$ &  $-$ &
 $-$ & $-$ & $2$ & $-$ &
 $-$ & $8$ & $-$ \\[0.5ex]
\hline\hline
\end{tabular}
\caption[]{The same as tab. (\ref{tab:2loop_plb}) for the $P$ diagrams
with a $\LR$ vertex insertion.}
\label{tab:2loop_prb}
\end{center}
\end{table}

\begin{table}
\begin{center}
\begin{tabular}{|c|cccc|}
\hline\hline
\multicolumn{1}{|c} { } &
\multicolumn{4}{|c|}{\tsq$\overline{D}$} \\
\hline
\multicolumn{1}{|c|}{$T_N$}&
$\frac{1}{\epdue}$ & $\tsq{\frac{1}{\epuno}}^{\sss HV}$ &
${\frac{1}{\epuno}}^{\sss NDR}$ & ${\frac{1}{\epuno}}^{\sss DRED}$ \\[0.5ex]
\hline
\tsp$P_2$ &
$-$ & $2$ & -$1$ & $2$ \\
\tsp$P_3$ &
$-$ & $2$ & $5$ & $2$ \\
\tsp$P_4$ &
$-$ & $-$ & $-6$ & $-$ \\
\tsp$P_{4}^{bg}$ &
$-$ & $-$ & $-5$ & $-$ \\
\tsp$P_5$ &
$-$ & $-$ & $-4$ & $-$ \\
\tsp$P_6$ &
$-$ & $-$ & $6$ & $-$ \\
\tsp$P_{6}^{bg}$ &
$-$ & $-$ & $5$ & $-$ \\
\tsp$P_7$ &
$-$ & $-$ & $-4$ & $-$ \\
\tsp$P_8$ &
$-$ & $-$ & $-1$ & $-$ \\
\tsp$P_9$ &
$-$ & $-$ & $-5$ & $-$ \\[0.5ex]
\hline\hline
\end{tabular}
\caption[]{The same as tab. (\ref{tab:2loop_plr}) for the $P$ diagrams
with a $\LR$ vertex insertion.}
\label{tab:2loop_prr}
\end{center}
\end{table}

\newpage
\clearpage

\begin{table}
\begin{center}
\begin{tabular}{|c|cccc|cccc|ccc|}
\hline\hline
\multicolumn{1}{|c} { } &
\multicolumn{4}{|c}{$D$} &
\multicolumn{4}{|c}{$C$} &
\multicolumn{3}{|c|}{$E$} \\
\hline
\multicolumn{1}{|c|}{$T_N$}&
$\frac{1}{\epdue}$ & $\tsq{\frac{1}{\epuno}}^{\sss HV}$ &
${\frac{1}{\epuno}}^{\sss NDR}$ &  ${\frac{1}{\epuno}}^{\sss DRED}$ &
$\frac{1}{\epdue}$ &
${\frac{1}{\epuno}}^{\sss HV}$ & ${\frac{1}{\epuno}}^{\sss NDR}$ &
 ${\frac{1}{\epuno}}^{\sss DRED}$ &
${\frac{1}{\epuno}}^{\sss HV}$ & ${\frac{1}{\epuno}}^{\sss NDR}$ &
${\frac{1}{\epuno}}^{\sss DRED}$ \\[0.5ex]
\hline
\tsp$F_2$ &
-$\frac{1}{9}$ & $\frac{71}{54}$ & $\frac{59}{54}$ & $\frac{25}{27}$ &
-$\frac{2}{9}$ & -$\frac{10}{27}$ & -$\frac{22}{27}$ & -$\frac{10}{27}$ &
$-$ & $-$ & -$\frac{7}{9}$ \\
\tsp$F_3$ &
$\frac{1}{9}$ & $\frac{91}{54}$ & $\frac{103}{54}$ & $\frac{29}{27}$ &
$\frac{2}{9}$ &  $\frac{10}{27}$ & $\frac{22}{27}$ & $\frac{10}{27}$ &
$-$ & $-$ & -$\frac{11}{9}$ \\
\tsp$F_4$ &
$\frac{1}{9}$ & $\frac{25}{54}$ & $\frac{37}{54}$ & $\frac{19}{54}$ &
$\frac{2}{9}$ & $\frac{10}{27}$ & $\frac{22}{27}$ & $\frac{10}{27}$ &
$-$ & $-$ & -$\frac{2}{9}$ \\
\tsp$F_{4}^{bg}$ &
$\frac{1}{9}$ & $\frac{25}{54}$ & $\frac{37}{54}$ & $\frac{19}{54}$ &
$\frac{2}{9}$ & $\frac{10}{27}$ & $\frac{22}{27}$ & $\frac{10}{27}$ &
$-$ & $-$ & -$\frac{2}{9}$ \\
\tsp$F_7$ &
$-$ & -$\frac{2}{9}$ & -$\frac{8}{9}$ & $\frac{1}{9}$ &
$-$ & $-$ & -$\frac{4}{3}$ & $-$ &
$-$ & $-$ & $\frac{2}{3}$ \\[0.5ex]
\hline\hline
\end{tabular}
\caption[]{The same as tab. (\ref{tab:2loop_plb}) for the $F$ diagrams
with a $\LR$ vertex insertion.}
\label{tab:2loop_frb}
\end{center}
\end{table}

\begin{table}
\begin{center}
\begin{tabular}{|c|cccc|}
\hline\hline
\multicolumn{1}{|c} { } &
\multicolumn{4}{|c|}{\tsq$\overline{D}$} \\
\hline
\multicolumn{1}{|c|}{$T_N$}&
$\frac{1}{\epdue}$ & $\tsq{\frac{1}{\epuno}}^{\sss HV}$ &
${\frac{1}{\epuno}}^{\sss NDR}$ & ${\frac{1}{\epuno}}^{\sss DRED}$ \\[0.5ex]
\hline
\tsp$F_2$ &
$\frac{1}{9}$ & $\frac{91}{54}$ & $\frac{103}{54}$ & $\frac{91}{54}$ \\
\tsp$F_3$ &
-$\frac{1}{9}$ & $\frac{71}{54}$ & $\frac{59}{54}$ & $\frac{71}{54}$ \\
\tsp$F_4$ &
-$\frac{1}{9}$ & $\frac{5}{54}$ & -$\frac{7}{54}$ & $\frac{5}{54}$ \\
\tsp$F_{4}^{bg}$ &
-$\frac{1}{9}$ & $\frac{5}{54}$ & -$\frac{7}{54}$ & $\frac{5}{54}$ \\
\tsp$F_7$ &
$-$ & -$\frac{2}{9}$ & $\frac{4}{9}$ & -$\frac{2}{9}$ \\[0.5ex]
\hline\hline
\end{tabular}
\caption[]{The same as tab. (\ref{tab:2loop_plr}) for the $F$ diagrams
with a $\LR$ vertex insertion.}
\label{tab:2loop_frr}
\end{center}
\end{table}

\newpage
\clearpage

\section{Anomalous Dimension Matrices}
\label{sec:dimanom}

The anomalous dimension matrix appearing in eq. (\ref{rge}) is defined
as
\be
\hat\gamma=\frac{\alphas}{4\pi}\left(
\hat\gamma_O+\frac{1}{2}\left(\gamma_f\hat n_f+\gamma_g\hat n_g
\right)-\gamma_{m_b}\hat S_1-\beta_0\hat S_2\right),
\label{gammao}
\ee
where the diagonal matrices $\hat n_f$ and $\hat n_g$ respectively count
the number of fermion and
gluon external fields of the operator they are applied to, $(\hat S_1)_{ij}=
\delta_{i7}\delta_{7j}+\delta_{i8}\delta_{8j}$, and
$(\hat S_2)_{ij}=\delta_{i8}\delta_{8j}$. The anomalous dimensions due
to the external fields and to the explicit couplings and masses are known to be
\bea
\gamma_f=2\frac{N^2-1}{2N} &,&
\gamma_{m_b}=-6\frac{N^2-1}{2N}\nn\\
\gamma_g=-2\left(\frac{11}{3}N-\frac{2}{3}n_f\right) &,&
\gamma_g^{bg}=-2\left(\frac{5}{3}N-\frac{2}{3}n_f\right)\nn\\
\beta(\alphas)=\frac{\alphas^2}{4\pi}\beta_0+\dots&=&
-\frac{\alphas^2}{4\pi}\left(\frac{11}{3}N-\frac{2}{3}n_f\right)+\dots
\label{gamext}
\eea
The two values of $\gamma_g$ refer to the Feynman and background gauge
calculations.

The operator anomalous dimension $\hat\gamma_O$ is defined
in terms of the matrix of the renormalization constants as shown in eq.
(\ref{diman}). In turn this matrix has an expansion, in terms of
the renormalized coupling constant $\alphas$ and the regularization
parameter $\ep$, more involved than in other cases.
In fact now the renormalization constants include an explicit dependence
on the subtraction scale $\mu$, starting already at order $O(\alphas^0)$,
due to a mismatch in the dimension between 4-fermion and magnetic operators.
Thus the usual expression of the anomalous dimension becomes
\be
\hat\gamma_O = 2\hat Z^{-1}\left[\left(-\epsilon\alphas+\beta(\alphas)
\right)\frac{\partial}{\partial\alphas}+\mu^2\frac{\partial}
{\partial\mu^2}\right]\hat Z,
\label{diman1}
\ee
while the multiple expansion of the matrix $\hat Z$ is given by
\bea
\hat Z&=&1+\mu^{-2\epsilon}\left(\hat Z^{0,1}_{0}+\hat Z^{0,1}_{1}\ep\right)+
\frac{\alphas}{4\pi}\left[\left(\hat Z^{1,1}_{0}+\hat Z^{1,1}_{1}
\ep\right)\right.\nn\\
&&+\left.\mu^{-2\epsilon}\left(\hat Z^{1,2}_{1}\ep+
\hat Z^{1,2}_{2}\frac{1}{\epdue}\right)\right]+\dots
\label{zexp}
\eea
The coefficients are labeled as $\hat Z^{a,b}_{c}$, where
$a$ is the order in $\alphas$, $b$ is the number of loops involved in the
calculation and $c$ is the order in the $\ep$ expansion.

Using eqs. (\ref{diman1}) and (\ref{zexp}), we obtain
\bea
\Z^{1,2}_2 &=& 0\nn\\
\hat\gamma_O &=& -2\hat Z^{1,1}_1-4\left(\hat Z^{1,2}_1-\frac{1}
{2}\hat Z^{1,1}_1
\hat Z^{0,1}_0-\frac{1}{2}\hat Z^{0,1}_1 \hat Z^{1,1}_0\right)
\label{expgam}
\eea
In order to have a finite $\hat\gamma_O$ when $D$ tends to 4 dimensions,
the first of eqs. (\ref{expgam}) must be satisfied.
Then the second equation gives the anomalous dimension matrix in terms of
the single pole coefficients and finite parts of the one and two loop diagrams,
see also eq. (\ref{rendiag}). This equation is obtained
by using the relations
\bea
\left(\hat Z^{0,1}_0\right)_{PP}&=&\left(\hat Z^{1,1}_0\right)_{PP}=
\left(\hat Z^{0,1}_1\right)_{PP}=0\nn\\
\left(\hat Z^{0,1}_0\right)_{PE}&=&\left(\hat Z^{1,1}_0\right)_{PE}=0\nn\\
\left(\hat Z^{0,1}_1\right)_{EP}&=&\left(\hat Z^{1,1}_1\right)_{EP}=
\left(\hat Z^{1,2}_1\right)_{EP}=0\nn\\
\hat Z^{0,1}_i\hat Z^{1,2}_j&=&
\hat Z^{1,2}_j\hat Z^{0,1}_i=0,\qquad i,j=0,1
\eea
which hold for our $\overline{MS}$ renormalization constants.
$P$ indices get values in the set of ``physical'', i.e. 4-dimensional,
operators, while $E$ refers to the ``effervescent'' ones.
Moreover only matrix elements between ``physical'' operators are retained,
the ``effervescent'' contribution being included in
the last two terms of $\hat\gamma_O$ in eqs. (\ref{expgam}).
In fact the summed indices in the products of the $\hat Z$ matrices
run over the full $D$-dimensional basis, including the ``effervescent''
operators.

The expression of $\hat\gamma_O$ in eq. (\ref{expgam}) is similar to the
next-to-leading order formula, see e.g. refs. \cite{cfmr2,bw1}, even if
this is a leading order calculation.
In particular the third term in eq. (\ref{expgam}) accounts for
the well-known ``effervescent'' contribution coming from one loop 4-fermion
diagrams. The fourth term includes the peculiar operators,
present only in the DRED scheme, which are generated in $4-D$ dimensions
by the diagrams in fig. (\ref{fig:1loop_ct}). Both these terms are actually
purely ``effervescent'', since they involve finite parts, which are not
included in the $\overline{MS}$ renormalization constants
as far as matrix elements between 4-dimensional operators are concerned.
On the contrary, matrix elements connecting ``physical'' and ``effervescent''
operators must be retained in the matrix of the renormalization
constants, even in the $\overline{MS}$ case.

In the last part of this section we summarize the results for the anomalous
dimension matrices in
the three considered schemes. The new calculation in the DRED scheme
gives a matrix which is identical to the HV
one. Hence the results reported here can also be found in ref.
\cite{cfmrs1}.

Splitting the anomalous dimension matrix as in eq. (\ref{matform}),
we give the results for $\hat\gamma_r$, $\vec\beta_7$, $\vec\beta_8$
and $\gamma_{77}$, $\gamma_{87}$, $\gamma_{88}$.

The regularization scheme independent matrix $\hat\gamma_r$ is
given by
\be
\hat{\gamma}_r=
\left( \begin{array}{cccccc}
-\frac{6}{N} & 6 &0 &0 &0 &0 \\
 & & & & & \\
6 &-\frac{6}{N} & -\frac{2}{3N} & \frac{2}{3} & -\frac{2}{3N} & \frac{2}{3} \\
 & & & & & \\
0 &0 &-\frac{22}{3N} &\frac{22}{3} &-\frac{4}{3N} &\frac{4}{3} \\
 & & & & & \\
0 &0 &6-\frac{2n_{\scriptscriptstyle f}}{3N}&-\frac{6}{N} + \frac{2
n_{\scriptscriptstyle f}}{3}&-\frac{2n_{\scriptscriptstyle f}}{3N} &\frac{2
n_{\scriptscriptstyle f}}{3}\\
 & & & & & \\
0 & 0 & 0 &0 &\frac{6}{N} &-6 \\
 & & & & & \\
0 &0 &-\frac{2n_{\scriptscriptstyle f}}{3N} & \frac{2n_{\scriptscriptstyle
f}} {3}  &-\frac{2n_{\scriptscriptstyle f}}{3N} &
-12\frac{N^{\scriptscriptstyle 2} -1}{2N} + \frac{2n_{\scriptscriptstyle
f}}{3}
\end{array} \right)
\label{gamma_r}
\ee
where $N$ is the number of colours and $n_f=n_u+n_d$ is the number of active
flavors.

The mixing of the magnetic operators $O_7$ and
$O_8$ among themselves is also scheme independent and is given by
\bea
\gamma_{77}&=&8\frac{N^{\scriptscriptstyle 2}-1}{2N} \nn \\
\gamma_{87}&=&8\frac{N^{\scriptscriptstyle 2}-1}{2N} \nn \\
\gamma_{88}&=&4N -\frac{8}{N}.
\label{gamma_mag}
\eea

For the vectors $\vec\beta$, which depend on the regularization scheme,
we obtain in HV and DRED
\bea
\vec{\beta}^{\scriptscriptstyle HV,\,DRED}_{7}&=&
\left( \begin{array}{c}
0 \\
\\
\frac{8}{9} \frac{N^{\scriptscriptstyle 2}-1}{2N} +
\frac{12Q_{\scriptscriptstyle
u}}{Q_{\scriptscriptstyle d}}\frac{N^{\scriptscriptstyle 2}-1}{2N} \\
\\
\frac{232}{9}\frac{N^{\scriptscriptstyle 2}-1}{2N} \\
\\
\frac{8n_{\scriptscriptstyle f}}{9}\frac{N^{\scriptscriptstyle 2}-1}{2N}
+\frac{12\bar{n}_{\scriptscriptstyle f}(N^{\scriptscriptstyle 2}-1)}{2N}\\
\\
-16\frac{N^{\scriptscriptstyle 2}-1}{2N} \\
\\
\frac{8n_{\scriptscriptstyle f}}{9}\frac{N^{\scriptscriptstyle 2}-1}{2N}
- \frac{12\bar{n}_{\scriptscriptstyle f}(N^{\scriptscriptstyle 2}-1)}{2N}
\end{array} \right)\nn\\
&&\\
\vec{\beta}^{\scriptscriptstyle HV,\,DRED}_{8}&=&
\left( \begin{array}{c}
6\\
\\
\frac{22N}{9} - \frac{58}{9N}\\
\\
\frac{44N}{9} - \frac{116}{9N} + 6 n_{\scriptscriptstyle f}\\
\\
12 + \frac{22Nn_{\scriptscriptstyle f}}{9}
- \frac{58n_{\scriptscriptstyle f}}{9N}\\
\\
-4N + \frac{8}{N} - 6 n_{\scriptscriptstyle f}\\
\\
 -8 -\frac{32Nn_{\scriptscriptstyle f}}{9}
+ \frac{50n_{\scriptscriptstyle f}}{9N}
\end{array} \right),\nn
\label{betahv}
\eea
where $\bar n_f=n_d+\frac{Q_u}{Q_d}n_u$. The NDR result is given by
\bea
\vec{\beta}^{\scriptscriptstyle NDR}_{7}&=&
\left( \begin{array}{c}
0\\
\\
\frac{-16}{9} \frac{N^{\scriptscriptstyle 2}-1}{2N} +
\frac{12Q_{\scriptscriptstyle
u}}{Q_{\scriptscriptstyle d}}\frac{N^{\scriptscriptstyle 2}-1}{2N}\\
\\
\frac{184}{9}\frac{N^{\scriptscriptstyle 2}-1}{2N}\\
\\
-\frac{16n_{\scriptscriptstyle f}}{9}\frac{N^{\scriptscriptstyle 2}-1}{2N}
+  \frac{12\bar{n}_{\scriptscriptstyle f}(N^{\scriptscriptstyle 2}-1)}{2N}\\
\\
40\frac{N^{\scriptscriptstyle 2}-1}{2N}\\
\\
-\frac{16n_{\scriptscriptstyle f}}{9}\frac{N^{\scriptscriptstyle 2}-1}{2N}
- \frac{12\bar{n}_{\scriptscriptstyle f}(N^{\scriptscriptstyle 2}-1)}{2N}
+\frac{40N(N^{\scriptscriptstyle 2}-1)}{2N}\\
\\
\end{array} \right)\nn\\
&&\\
\vec{\beta}^{\scriptscriptstyle NDR}_{8}&=&
\left( \begin{array}{c}
6 \\
\\
\frac{22N}{9} - \frac{46}{9N}\\
\\
\frac{44N}{9} -\frac{92}{9N} + 6 n_{\scriptscriptstyle f}\\
\\
12 + \frac{22Nn_{\scriptscriptstyle f}}{9}
- \frac{46n_{\scriptscriptstyle f}}{9N}\\
\\
4N - \frac{20}{N} - 6 n_{\scriptscriptstyle f}\\
\\
-8 -\frac{32Nn_{\scriptscriptstyle f}}{9}
+ \frac{62n_{\scriptscriptstyle f}}{9N}
\end{array} \right).\nn
\label{betandr}
\eea

\section{Status of the Calculation of the QCD Correction to the
$b \rightarrow s \, \gamma$ Decay}
\label{sec:status}

Let us start with the original calculations, from ref. \cite{gsw1} to
ref. \cite{ccrv}. These works used the ``reduced'' basis. As we have shown,
this approximation leads to scheme dependent results. Apart from this,
there is no computational error in both NDR and DRED calculations.
However refs. \cite{gsw1,ccrv} did not include the contribution coming from
the $O(\alphas^0)$ mixing among $Q_5,\, Q_6$ and the magnetic operators,
which is present in NDR. On the other hand this mixing vanishes in DRED.
However the authors of ref. \cite{gosn1} overlooked the contributions of
the ``effervescent'' counter-terms. These two calculations obtained
different final results. This difference led the authors of ref.
\cite{gosn3} to the incorrect conclusion that the DRED scheme fails in this
case. Now we know that those results can differ (and indeed
they do) without implying any failure of the regularization scheme.
A second, more specific, argument in ref. \cite{gosn3} was based on the
explicit calculation of the sum of the diagrams $P2+P3$ in fig.
(\ref{fig:2loop}). They computed this sum in DRED and in a 4-dimensional
regularization scheme and obtained again different results.
Incidentally eqs. (\ref{check3}) show that the sum $P2+P3$ is indeed scheme
independent. We have checked that the two results of ref. \cite{gosn3}
actually coincide, once one includes the contribution of the DRED
``effervescent'' counter-terms.

Coming to more recent calculations, the full basis and the already mentioned
$O(\alphas^0)$ mixing have been taken into account, starting from ref.
\cite{mis1}, so that now problems with the scheme independence of the Effective
Hamiltonian are no more present, as shown in ref. \cite{cfmrs1}. However
the three latest works on the subject \cite{mis2,yao1,cfmrs1} give three
different
results for some anomalous dimension matrix elements. In particular,
using our normalization,
\begin{itemize}
\item ref. \cite{mis2} gives $\gamma_{57}=-32$, $\gamma_{67}=\frac{4432}{27}$,
$\gamma_{58}=10$, $\gamma_{68}=-\frac{2210}{27}$;
\item ref. \cite{yao1} gives $\gamma_{57}=\frac{416}{3}$, $\gamma_{67}=
\frac{7888}{27}$, $\gamma_{58}=-\frac{106}{3}$, $\gamma_{68}=-\frac{914}{27}$;
\item ref. \cite{cfmrs1} gives $\gamma_{57}=\frac{160}{3}$, $\gamma_{67}=
\frac{4432}{27}$, $\gamma_{58}=-\frac{74}{3}$, $\gamma_{68}=-\frac{1346}{27}$;
\end{itemize}
when $N=3$ and $n_f=5$.
The origins of these differences have been clearly summarized in ref.
\cite{mis3}. We shortly repeat them here, taking our calculation
as a reference
\begin{itemize}
\item the calculation of ref. \cite{mis2} differs from our one because
of the values of the diagrams $P2$ and $P3$ in tab. (\ref{tab:2loop_prr});
\item the calculation of ref. \cite{yao1} differs from our one because
the results presented there do not include the ``effervescent'' counter-terms,
as also stated in ref. \cite{yao2}.
\end{itemize}
Of course refs. \cite{mis2} and \cite{yao1} differ one from the other for
both the two reasons listed above.

In our opinion, the results of ref. \cite{yao1} actually agree with our
ones. In fact we cannot see
any reason not to include the ``effervescent'' counter-terms, which are known
to be needed from a long time\footnote{Incidentally the final result is not
scheme independent without the ``effervescent'' counter-terms.}, see also
eq. (\ref{expgam}). Concerning
ref. \cite{mis2}, we can just say that our diagrams, including those ones
responsible for the difference, verify all the checks we have done, including
those enforced by the scheme independence of the Effective Hamiltonian, eqs.
(\ref{check3}). Up to now we have not been able to find any
error in the calculations of $P2$ and $P3$ with a $m_b$ mass
insertion into the loop, which are actually quite easy to evaluate. Thus
we are confident that our results are correct.

Anyway these differences still present in the literature are known to have a
very little impact on the phenomenology of the radiative $B$ decays. We plan to
present our phenomenological analysis in a forthcoming paper \cite{cfmrs2}.

\section*{Acknowledgments}
We are grateful to G. Martinelli for his precious support in terms of
suggestions and discussions.
We acknowledge the partial support of the MURST, Italy, and INFN.

\end{document}